\documentclass[prl,twocolumn,showpacs,aps,floatfix]{revtex4}
\usepackage{graphicx}

\begin{document}

\title{Ferromagnetism and Kondo Insulator Behavior in the \\
Disordered Periodic Anderson Model}

\author{Unjong Yu$^1$}
\author{Krzysztof Byczuk$^{1,2}$}
\author{Dieter Vollhardt$^1$}

\affiliation{$^1$Theoretical Physics III, Center for Electronic
  Correlations and Magnetism, Institute for Physics, University of
  Augsburg, D-86135 Augsburg, Germany\\
  $^2$Institute of Theoretical Physics,
  University of Warsaw, ul. Ho\.za 69, 00-681 Warszawa, Poland
}

\date{\today}

\begin{abstract}
The effect of binary alloy disorder on the ferromagnetic phases of
$f$-electron materials is studied within the periodic Anderson
model. We find that disorder in the conduction band can drastically
enhance the Curie temperature $T_c$ due to an increase of the local
$f$-moment. The effect may be explained qualitatively and even
quantitatively by a simple theoretical \emph{ansatz}. The emergence
of an alloy Kondo insulator at non-integer filling is also pointed
out.
\end{abstract}

\pacs{71.23.-k, 75.20.Hr, 75.30.Mb}

\maketitle

Materials with $f$-electrons such as the rare earths (e.g., cerium)
or actinides (e.g., uranium) exhibit a wealth of highly unusual
thermodynamic, magnetic and transport properties
\cite{f_electron+HF}. The minimal microscopic model that can account
for this diverse physical behavior is the periodic Anderson model
(PAM) which describes a band of non-interacting electrons
hybridizing with localized, interacting $f$-electrons \cite{Lee86}.
Depending on the position of the $f$-electron energy $\varepsilon^f$
relative to the conducting band, and on the strength of the
hybridization $V$ and the local Coulomb interaction $U$, the PAM is
able to reproduce heavy fermion, intermediate valence, and local
moment behavior.
In the local moment regime and for large $U$ the PAM reduces to the
so called Kondo lattice model, which may be employed as an effective
model for manganites exhibiting colossal magnetoresistance
\cite{Dagotto01}, or for diluted magnetic semiconductors which show
promise for applications in spintronics \cite{DMS}. At low enough
temperatures the PAM also describes magnetically ordered phases.
While antiferromagnetic order is well-known to occur close to
half-filling \cite{Lee86}, ferromagnetic solutions are found far
away from half-filling \cite{FM_in_PAM}.
Indeed, ferromagnetism has been observed in various $f$-compounds
\cite{Durakiewicz04}.

Alloys of $f$-electron materials also display intriguing properties.
For example, by changing the stoichiometric composition of
Ce(Pt$_{1-x}$Ni$_{x}$)$_{2}$Si$_{2}$ the systems can be tuned from
the local moment regime at $x=0$ to the intermediate valence regime
at $x=1$ \cite{Ragel04}. Alloying inevitably introduces
\emph{disorder} into the system. In general, disorder is expected to
reduce the tendency towards ferromagnetic long-range order of the
$f$-electrons and thus lower the Curie temperature $T_c$. On the
other hand, in certain cases disorder is even known to improve the
stability of ferromagnetism. For example, disorder in the conduction
electron band caused by the substitution of Rh by Co in
URh$_{1-x}$Co$_x$Ge leads to a maximum in $T_c$ at $x\approx 0.6$
\cite{Sakarya07}. Similarly, a maximum in $T_c$ is observed in
CeCu$_2$Si$_{2-x}$Ge$_x$ at $x\approx 1.5$ which may be attributed
to an enhanced exchange interaction between the $f$-electron moments
induced by the diffusive motion of the Cu electrons \cite{Neto03}.
Finally, alloying Ce with La as in
(Ce$_{1-x}$La$_x$)$_3$Bi$_4$Pt$_3$ introduces disorder into the
$f$-electron system, which can trigger a transition from a Kondo
insulator to a dirty metal \cite{Hundley91}. Clearly disorder is an
important feature of many $f$-electron alloys and must therefore be
included in any comprehensive theoretical study of such compounds.

Previous investigations of the disordered PAM focussed, in
particular, on the effect of nonmagnetic impurities on the heavy
Fermi liquid or the Kondo insulating state \cite{Kondo-holes}, and
on the disorder-driven non-Fermi liquid behavior in Kondo alloys
\cite{Miranda97}. Grenzebach {\it et al.} \cite{Grenzebach07}
recently presented a detailed study of transport properties of the
disordered PAM within the dynamical mean-field theory (DMFT)
\cite{Georges96} together with a thorough discussion of the
development of the field. The effect of disorder in the
$f$-electrons on the ferromagnetic phase was investigated by Meyer
\cite{Meyer02}, who found that the Curie temperature is always
reduced.

In this Letter we report results of a detailed study of
ferromagnetism in the PAM in the presence of alloy disorder. In
particular, we show that $T_c$ can be substantially enhanced by
disorder in the \emph{conduction electrons}. We also predict Kondo
insulator behavior away from half-filling at particular values of
the alloy concentration. Quite generally, the Hamiltonian of the PAM
in the presence of local disorder has the form
\begin{eqnarray}
H &=& \sum_{ i,j \sigma} t_{ij} c^{\dagger}_{i \sigma}
c^{}_{j \sigma}
 + \sum_{i \sigma} \left( \varepsilon_{i}^f f^{\dagger}_{i \sigma}
f^{}_{i \sigma} + \varepsilon_{i}^c c^{\dagger}_{i \sigma}
c^{}_{i \sigma} \right)
  \nonumber \\ &&
+  \sum_{i \sigma} \left( Vc^{\dagger}_{i \sigma}
f^{}_{i \sigma} + V^*f^{\dagger}_{i \sigma}c^{}_{i \sigma} \right) + U
\sum_{i} n^{f}_{i \uparrow} n^{f}_{i \downarrow},
\label{hamiltonian}
\end{eqnarray}
where $c^{\dagger}_{i \sigma}$ ($c_{i \sigma}$) and
$f^{\dagger}_{i\sigma}$ ($f_{i\sigma}$) are creation (annihilation)
operators of conduction ($c$) and localized ($f$) electrons with
spin $\sigma$ at a lattice site $i$.
Here the on-site energies
$\varepsilon_{i}^f$ and $\varepsilon_{i}^c$ are random
variables and $V$ is the local hybridization between $f$- and
$c$-electrons.
The hopping amplitude of the $c$-electrons is given by $t_{ij}$. The
Coulomb interaction $U$ acts only between $f$-electrons on the same
site.

The alloy will be modeled by a bimodal probability distribution
function, $ P(y_i) =  x \delta(y_i-y_{0}+\Delta^y) +  (1-x)
\delta(y_i-y_{0})$, where $y_i=\varepsilon^c_i$, $\varepsilon_i^f$
are independent, random variables with reference values
$y_0=\varepsilon^c_0$, $\varepsilon^f_0$. The alloy concentration is
characterized by the parameter $x$ and the splitting between the
atomic levels of the alloy components by the energy
$\Delta^{y}=\Delta^c$, $\Delta^f$, respectively. While the
concentration $x$ and energy splitting $\Delta^y$ are, in general,
independent parameters the values $x=0$,~$1$ correspond to a
non-disordered system even if $\Delta^y \ne 0$. Hence
$\delta^y=x(1-x)\Delta^y$ is a natural parameter for the disorder
strength of alloy disorder.

The PAM with binary alloy disorder is solved within DMFT \cite{Georges96},
which becomes exact in the limit of infinite dimensions \cite{Metzner89}.
In DMFT the disordered PAM is mapped onto independent impurities, i.e.,
for each random variable $\{y_i\}$ the action has the form
\begin{widetext}
\begin{eqnarray}
S^{\rm loc}[f_{\sigma},c_{\sigma};\{y_i\}] =
\sum_{\sigma n}\left(
  f^*_{\sigma n},c^*_{\sigma n}\right)
\left(
\begin{array}{cc}
i\omega_n+\mu -\varepsilon^f_i&V^*\\
V&i\omega_n+\mu-\varepsilon_i^c-\eta_{\sigma n}
\end{array}
\right) \left(
\begin{array}{c}
f_{\sigma n}\\
c_{\sigma n}
\end{array}
\right)
+ U\int_0^{\beta}\; d\tau\; n^f_{\uparrow}(\tau)\; n^f_{\downarrow}(\tau).&
\label{action}
\end{eqnarray}
\end{widetext}
The function
$\eta_{\sigma n}$ describes an effective dynamical hybridization
of the $c$-electrons with the bath. It is the same for all random
variables $\{y_i\}$ and is determined by the self-consistency
equations to be discussed below. We start with the local matrix
Green function
\begin{eqnarray}
{\bf G}_{\sigma}^{\rm loc}(\tau;\{y_i\}) = - \left( \!
\begin{array}{cc}
\langle T_{\tau} f^{}_{\sigma}(\tau) f_{\sigma}^{\dagger}(0)\rangle &
\langle T_{\tau} f^{}_{\sigma}(\tau) c_{\sigma}^{\dagger}(0)\rangle \\
\langle T_{\tau} c^{}_{\sigma}(\tau) f_{\sigma}^{\dagger}(0)\rangle &
\langle T_{\tau} c^{}_{\sigma}(\tau) c_{\sigma}^{\dagger}(0)\rangle
\end{array}
\! \right),
\label{local_green}
\end{eqnarray}
where $T_{\tau}$ is the time-ordering operator.  Since the Green
function (\ref{local_green}) depends on $\{y_i\}$ it is a random
function.
Here we perform arithmetic averaging,
i.e., the averaged Green function is given by
\begin{eqnarray}
{\bf G}_{\sigma }^{\rm loc}(\tau)= \int \prod_{\{y_i\} }d y_i
P(y_i) {\bf G}_{\sigma}^{\rm loc}(\tau;\{y_i\}).
\label{averaged_green}
\end{eqnarray}
In the absence of interactions one then obtains the results of the
well-known coherent potential approximation (CPA) \cite{Vlaming92}.
Effects of Anderson localization  are neglected in this case but can
be incorporated by employing the geometric average \cite{Byczuk05}.
The self-consistency requires the averaged local Green function
(\ref{averaged_green}) to be the same as the lattice Green function,
i.e.,
\begin{eqnarray}
{\bf G}_{\sigma n} = \sum_{\bf k} \left(
\begin{array}{cc}
i\omega_n+\mu -\Sigma^f_{\sigma n} & V^*  \\
V  & i\omega_n+\mu-\epsilon_{\bf k} -
\Sigma^c_{\sigma n}
\end{array}
\right)^{-1}.
\label{lattice_green}
\end{eqnarray}
The local self-energies appear in the $\bf k$-integrated Dyson
equation ${\bf \Sigma}_{\sigma n}={\bf \cal G}^{-1}_{\sigma n} -
{\bf G}_{\sigma n}$, where ${\bf \cal G}_{\sigma n}$ is the local
Green function of the non-interacting bath electrons, with
\begin{eqnarray}
{\bf \cal G}_{\sigma n}^{-1}=\left(
\begin{array}{cc}
i\omega_n+\mu-\varepsilon^f_0&V^*\\
V&i\omega_n+\mu-\varepsilon^c_0-\eta_{\sigma n}
\end{array}
\right). \label{weiss_green}
\end{eqnarray}
Eqs.~(\ref{action}-\ref{weiss_green}) form a general, closed set of
equations, which determine all local, dynamical correlation functions
of the disordered PAM.

To understand the effect of alloy disorder on the physics described
by the PAM  it is instructive to investigate the case $U=0$ first,
since alloy disorder affects a hybridized two-band system in several
interesting ways. To this end we consider the disorder to act only
on the $c$-electrons or the $f$-electrons, respectively. In the case
of $c$-electron disorder the diagonal elements of the local Green
function (\ref{lattice_green}) are given by
\begin{eqnarray}
G^{cc}_{\sigma n}&\!\!=\!\!&
   \frac{x}{({\cal G}^{cc}_{\sigma n})^{-1} -|V|^2
     {\cal G}^{ff}_{\sigma n}}
   + \frac{1-x}{({\cal G}^{cc}_{\sigma n})^{-1}
     -|V|^2 {\cal G}^{ff}_{\sigma n} - \Delta^c} \nonumber \\
G^{ff}_{\sigma n} &\!\!=\!\!&
   \frac{x}{({\cal G}^{ff}_{\sigma n})^{-1} -
     |V|^2 {\cal G}^{cc}_{\sigma n} }
   +\frac{1-x}{({\cal G}^{ff}_{\sigma n})^{-1} -
     \frac{|V|^2}{({\cal G}^{cc}_{\sigma n})^{-1} - \Delta^c }}.
\label{diagonal}
\end{eqnarray}
The case of $f$-electron disorder  is obtained by exchanging
$f\leftrightarrow c$ in Eq.~(\ref{diagonal}). Large energy splitting
$\Delta^c$ leads to a band splitting of the conduction electrons as
in the single band model \cite{Byczuk04}, i.e., each alloy subband
contains $2xN_L$ and $2(1-x)N_L$ states, respectively, where $N_L$
is the number of lattice sites. At the same time, the $f$-level does
not split. Altogether the alloy with hybridized
$c$- and $f$-electrons can therefore be a band insulator even for
total densities different from integer values ($2$ or $4$)
\cite{Byczuk04}.

We now include the interaction $U$ between the $f$-electrons and
investigate its influence on the alloy subbands. The effective
two-orbital impurity problem in the presence of disorder,
Eqs.~(\ref{action}-\ref{weiss_green}), is solved by finite
temperature determinant quantum Monte-Carlo (QMC) simulations.
Ferromagnetic instabilities are detected by the divergence of the
static spin susceptibility and by a non-vanishing value of the
magnetization \cite{Ulmke98+Byczuk02}. In the numerical examples
presented below the DOS for the non-interacting $c$-electrons has
the model form $N_0(\varepsilon) =\sqrt{4-\varepsilon^2}/2\pi$,
where the energy unit is $t=1$. In the following we fix the
interaction and the hybridization at $U=1.5$ and $V=0.6$,
respectively, and include disorder either in the $f$-electron  or
$c$-electron system.

\begin{figure}[tbp]
\includegraphics[width=8.2cm]{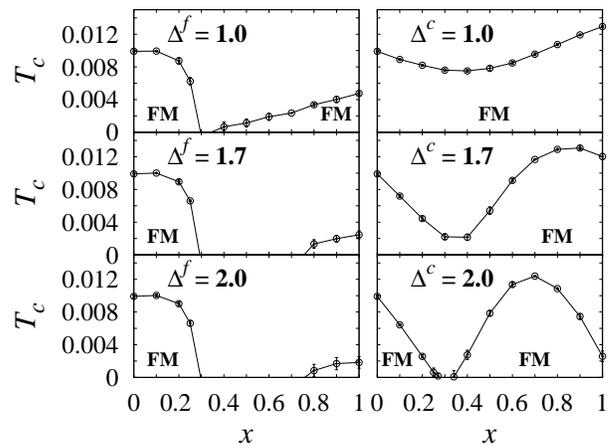}
\caption{\label{fig:TcD} Curie temperature $T_c$ as a function of
alloy concentration $x$ and energy splitting $\Delta^f$ (left
column) and  $\Delta^c$ (right column) for $n_{\rm tot}=1.3$ and
$\varepsilon_c^0-\varepsilon_f^0=3.25$. Strong $c$-electron disorder
enhances $T_c$ compared to its values at $x=0$ or $1$.}
\end{figure}

As shown in Fig.~\ref{fig:TcD} the computed Curie temperature for
the transition to the ferromagnetic state is a non-monotonic
function of the alloy concentration $x$. In particular, the behavior
is quite different for disorder acting on the $f$- or the
$c$-electrons.

\underline{$f$-electron disorder:}
 In agreement with Meyer \cite{Meyer02} the presence of
$f$-electron disorder always reduces the Curie temperature
relative to its non-disordered values at $x=0$ or 1.
For strong enough disorder $T_c$ eventually vanishes, e.g., at
$x=0.28$ and $x=0.75$, respectively, for $\Delta^f=1.7$ (left panel of
Fig.~\ref{fig:TcD}).
This is due to the splitting of the $f$-electron band at large
$\Delta^f$ which increases the double occupation of the lower alloy
subband; this reduces the local moment of the $f$-electrons and
thereby $T_c$.

\underline{$c$-electron disorder:} By contrast, $c$-electron
disorder leads to a much more subtle dependence of $T_c$ on
concentration $x$. Namely, for increasing energy splitting
$\Delta^c$ there are, in general, three different features observed,
 the physical origin of which will be discussed in more detail
later: (i) at $x=1$, i.e., in the non-disordered case, $T_c$ is
reduced; (ii) a minimum develops in $T_c$ at $x=n_{\rm tot}-1>0$;
(iii)  $T_c$ is \emph{enhanced} over its non-disordered values at
$x=0$ or $1$. Altogether this leads to a global maximum in $T_c$ vs.
$x$. While the decrease of $T_c$ at $x=1 $ is a simple consequence
of the reduction of the energy difference between the $f$-level and
the $c$-electron band, $\varepsilon^c - \varepsilon^f =
\varepsilon^c_0 - \varepsilon^f_0 - \Delta^c$, for increasing
$\Delta^c$, the latter effects are more subtle.

\begin{figure}[tbp]
\includegraphics[width=7cm]{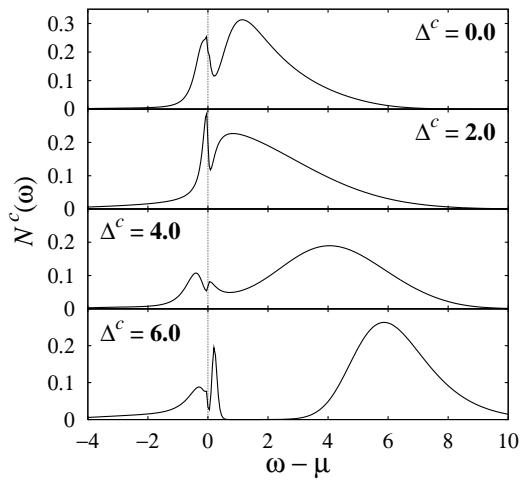}
\caption{\label{fig:Awc} Spectral function of $c$-electrons for
different $\Delta^c$ at $x=0.3$ (other parameters as in
Fig.~\protect\ref{fig:TcD}) obtained within QMC and maximal entropy
at $T=1/60$. By increasing $\Delta^c$ a pseudogap opens which
becomes a real gap for $T \rightarrow 0$. }
\end{figure}

To understand the minimum in $T_c$ vs. $x$  we computed the
evolution of the spectral functions $N^c(\omega)$ at $x=0.3$ for
$\Delta^c$ increasing from $0$ to $6$ (Fig.~\ref{fig:Awc}). There is
an opening of a gap at the chemical potential signalling a
metal-insulator transition in this system. This is caused by the
splitting of the $c$-electron band due to binary alloy disorder and
the correlations between the $f$-electrons. Namely, for energy
splittings $\Delta^c$ much larger than the width of the $c$-electron
band the total number of available low-energy states is reduced from
$4 N_L$ to $[4 - 2(1-x)]N_L = 2(1+x)N_L$, whereby the filling
effectively increases by a factor of $4/[2(1+x)]$, such that $n_{\rm
tot}^{\rm eff}=2 n_{\rm tot}/(1+x)$, if $n_{\rm tot}<2(1+x)$. For
the filling $n_{\rm tot}=1.3$ studied in the
Figs.~\ref{fig:TcD}~and~\ref{fig:Awc}  the concentration $x=0.3$ is
a special case since then $n_{\rm tot}^{\rm eff}=2$. The system is
then effectively at half-filling and behaves as a Kondo insulator at
large $U$, $\Delta^c$,  and low temperatures. In particular,
itinerant ferromagnetism is unfavorable in this case, i.e., $T_c=0$
in the vicinity of $x=0.3$ at $\Delta^c=2$, cf. Fig.~\ref{fig:TcD}.
The metal to Kondo insulator transition at non-integer filling in
the PAM predicted here is a counterpart of the Mott-Hubbard metal to
insulator transition at non-integral fillings in the one-band
Hubbard model found in \cite{Byczuk03,Byczuk04}.

\begin{figure}[tbp]
\includegraphics[width=8.2cm]{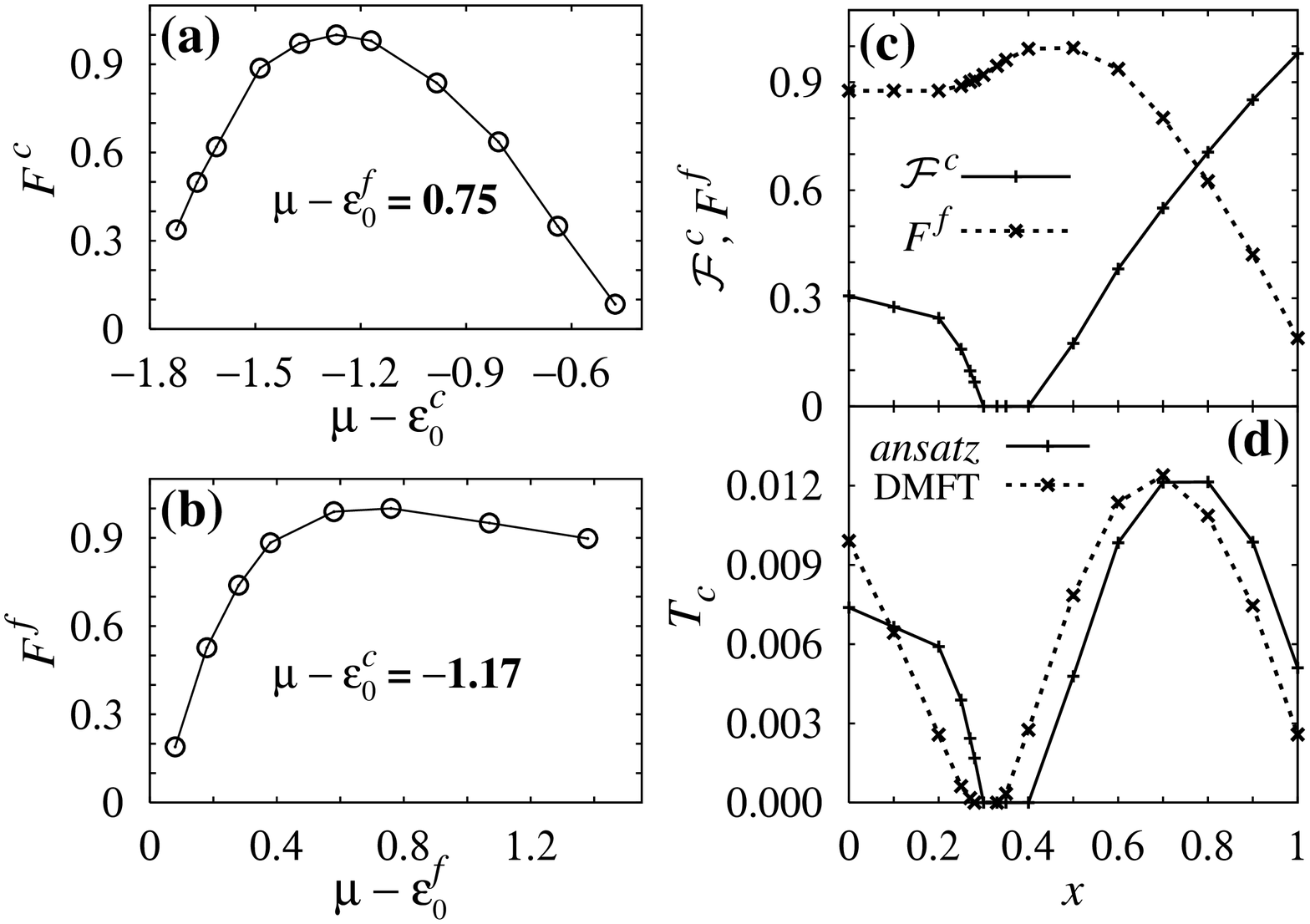}
\caption{\label{fig:scaling} (a) $F^c(\mu-\varepsilon^c_0)$,  (b)
$F^f(\mu-\varepsilon^f_0)$ appearing in the \emph{ansatz} for $T_c$
(see text) calculated for $\Delta^c = 0$. (c) ${\cal
F}^{c}(\mu-\varepsilon^c_0)$ and $F^f(\mu-\varepsilon^f_0)$ for
$\Delta^c = 2.0$; other parameters as in Fig.~\protect\ref{fig:TcD}.
(d) Comparison of $T_c$ obtained from the \emph{ansatz} and within
DMFT.}
\end{figure}

We now turn to the maximum in $T_c$ vs. $x$.
It can be understood within the following
model based on an \emph{ansatz} for the Curie temperature,
$T_c(U,V,\mu)=T_c^{0}(U,V,\mu)F^c(\mu-\varepsilon_c^0)
F^f(\mu-\varepsilon_f^0)$, which implies that the formation of local
$f$-electron moments ($F^f$) is assumed to be independent from the
$c$-electron mediated ordering of those moments ($F^c$). In
fact, for the RKKY model this \emph{ansatz} can be microscopically
justified within a static mean-field theory \cite{remark}.  The two
functions $F^{c}$, $F^f$ are determined by $T_c$ calculated within
DMFT for the non-disorder case at fixed $\mu-\epsilon^c_0$ or
$\mu-\epsilon^f_0$, respectively; they are shown in Fig.~\ref{fig:scaling}(a) and
\ref{fig:scaling}(b) for one set of parameters. The prefactor $T_c^0$
is determined
by the requirement that the dimensionless functions $F^f$ and $F^c$
be equal to one at their maxima. We note that
$F^f(\mu-\varepsilon^f_0)$ has a maximum when the $f$-level is
half-filled ($\mu=\varepsilon^f_0+U/2$), i.e., when the local moment
is maximal.

The Curie temperature in the presence of $c$-electron disorder can
now be estimated by averaging over the $c$-electron part, $F^c$,
giving rise to the disorder-dependent function ${\cal
F}^c(x,\mu-\epsilon^c_0)=[x F^c(\mu-\varepsilon^c_0+\Delta^c) +(1-x)
F^c(\mu-\varepsilon^c_0)]$. The linear dependence on the
alloy concentration can again be justified microscopically within a
static mean-field theory for the RKKY model, where $T_c$ depends
linearly on the DOS at the chemical potential \cite{remark}. $T_c$
is now determined for each concentration $x$. We calculate $\mu$,
which is  an implicit function of $x$, in the non-hybridized limit
($V=0$) within a rigid band approximation \cite{remark2}. The
dependence of the resulting functions ${\cal
F}^{c}(x,\mu-\varepsilon^c_0)$ and $F^f(\mu-\varepsilon^f_0)$ on $x$
are shown in Fig.~\ref{fig:scaling}(c) for $\Delta^c = 2.0$.  In
general $F^f(\mu-\epsilon^f_0)$ has a global maximum at those values
of $x$ for which the $f$-level is half-filled [see
Fig.~\ref{fig:scaling}(c)]. By contrast, ${\cal
F}^c(x,\mu-\epsilon^c_0)$ is characterized by a wide minimum,
related to the formation of the pseudo-gap in the interacting DOS
seen in Fig.~\ref{fig:Awc}. This minimum reaches zero, i.e., ${\cal
F}^c(x,\mu-\varepsilon^c_0)=0$, for a finite range of $x$ values as
 shown in Fig.~\ref{fig:scaling}(c). The resulting $T_c(x)$
obtained by the product of these two functions agrees remarkably
well with the numerical result obtained by DMFT as shown in
Fig.~\ref{fig:scaling}(d).

In conclusion, the interplay between the disorder induced splitting
of the conduction band and many-body correlations among the
$f$-electrons can lead to a remarkable enhancement of the Curie
temperature in the periodic Anderson model. There are two competing
effects determining $T_c$ as the alloy concentration $x$ is
decreased from $x=1$: (i) a rise due to an increase of the local
moment, and (ii) a decrease due to the opening of a gap in the alloy
Kondo insulator at non-integer filling. Altogether this leads to a
global maximum in $T_c$ vs. $x$. Therefore experimental
investigations of $f$-electron materials with alloy disorder in the
conducting band are expected to be particularly rewarding.

This work was supported in
part by the Sonderforschungsbereich 484 of the Deutsche
Forschungsgemeinschaft (DFG).


\begin{thebibliography}{99}
\bibitem{remark2} Approximating the chemical
  potential by
  $\mu=x\mu_{x=1}+(1-x)\mu_{x=0}$, leads to qualitatively
  similar results as long as $\Delta^c$ is smaller than the band width.
\end{thebibliography}
\end{document}